\documentclass[prb,superscriptaddress,reprint,amssymb,aps,floatfix,showkeys]{revtex4-1}

\usepackage{amsmath}%
\usepackage{amsfonts}%
\usepackage{amssymb}%
\usepackage{graphicx}
\usepackage{color}
\usepackage{tabularx}
\usepackage{braket}

\begin{document}

\title{Spin bath maser in a cryogenically cooled sapphire whispering gallery mode resonator}
	
\author{J. Bourhill}
\author{K. Benmessai}
\author{M. Goryachev}
\email{maxim.goryachev@uwa.edu.au}

\author{D.L. Creedon}
\author{W. Farr}
\author{M.E. Tobar}	

\affiliation{ARC Centre of Excellence for Engineered Quantum Systems, University of Western Australia, 35 Stirling Highway, Crawley WA 6009, Australia}

\date{\today}
\begin{abstract}
We report the observation of a mechanism of maser generation in an ensemble of inter-coupled, inhomogeneously broadened two-level systems, enhanced by high quality factor electromagnetic cavity modes. In this previously unobserved form of population inversion, an inseparable quantum system leads to cavity-enhanced stimulated emission arising from interactions within an ensemble of two-level systems, as opposed to a traditional ensemble of noninteracting identical three level systems.
The effect is observed in a cryogenically cooled whispering gallery mode sapphire resonator containing dilute Fe$^{3+}$ impurity ions. 
These ions exhibit strong spin-lattice interaction, leading to both electron spin resonance broadening and phonon mediated spin-spin coupling. The maser effect is due to a $\left|1/2\right\rangle \rightarrow \left|3/2\right\rangle$ energy transition in electron spin angular momentum observed at zero external magnetic field. Both continuous and oscillating regimes are observed with corresponding thresholds both in detuning frequency and incident power.
 
\end{abstract}
\maketitle

The fundamental purpose of a maser system is to achieve a net microwave emission from a solid or gas, which is reduced to a problem of populating an upper energy level in excess compared to lower levels. There are a number of successful solutions to this problem, for instance, the utilisation of three-level schemes in solid state systems, state selection of particles in gaseous masers, pulsed techniques, and the application of an external magnetic field. 
Zero field solid-state masers have been investigated since the late 1950s\cite{{50},{60}}, and operate due to the presence of residual or intentionally doped paramagnetic ion impurities within a crystal lattice.
These noninteracting ions exhibit an Electron Spin Resonance (ESR), typically forming multiple-level systems at zero applied DC magnetic field due to the strong electric field of the host crystal which acts to to split the electron spin energy levels \cite{{karim1},{karim2},{maser1},{maser2},{maser3},{maser4},{maser5}}. In the traditional approach, some of these energy levels are used to implement a three-level scheme, where the required population inversion is achieved by externally pumping electrons from the ground state (e.g. $\Ket{1/2}$) to an upper energy state (e.g. $\Ket{5/2}$) from where they eventually (non-radiatively) relax into the intermediate state $\Ket{3/2}$. At a certain threshold pump power, the intermediate and ground states constitute a two-level system with the inverted population suitable for amplification by stimulated emission. The efficiency of this approach has been recently improved by utilising high $Q$-factor Whispering Gallery (WG) modes in cylindrical crystals\cite{karim2} where high ion-photon interaction probabilities are achieved via resonant confinement of photons.

In this work, we demonstrate a new scheme resulting in population inversion of impurity ions in a solid state crystal, and consequently a net microwave emission. The novel approach utilises several favourable properties of sapphire WG mode resonators and ESRs of Fe$^{3+}$ ions: high $Q$-factors of the WG modes, their relative abundance, inhomogeneous broadening of the ESR, and weak spin-spin interaction of the ensemble of spins. 

The solid state crystal used to implement the system is a highest-purity cylindrical sapphire crystal\cite{fwm} grown using the heat exchange method by GT Crystal Systems (USA) with dimensions 30 mm height $\times$ 50 mm diameter. The crystal is mounted inside a cylindrical copper cavity with silver-plated internal walls, with the entire assembly held under vacuum and cooled to $\sim 4.2$~K. Fe$^{3+}$ ions substitute Al$^{3+}$ cations in the sapphire lattice as residual impurities during the manufacturing process. These ions have have a $^6$S ground state with an electronic configuration of the unfilled shell 3d$^5$. Zero field splitting (into three energy levels) of this ground state is due to the strong internal crystal field of the sapphire lattice\cite{KorPro}. The Zeeman effect splits these three levels into two sub levels each in the presence of a DC magnetic field, giving rise to six quantum energy states of the ion: $\Ket{\pm1/2}$, $\Ket{\pm3/2}$, and $\Ket{\pm5/2}$. Due to a post-growth annealing of the crystal, such impurities exist in higher concentrations than values reported in the literature for other high purity sapphire \cite{{crystal1},{crystal2}}, with a concentration in the present case of 150 parts per billion\cite{maser5}, corresponding to $\sim 4 \times 10^{21}$ ions/m$^3$.\\

For a crystal of these dimensions, a quasi-transverse magnetic WG mode (WGH) of 17$^{\text{th}}$ azimuthal order with no axial or radial nodes exists at $\sim$12.038~GHz, whereas the `signal' modes near 11.8~GHz correspond to lower $Q$-factor modes of different mode families. The $Q$-factors of these modes are on the order of $10^{9}$ and $10^7$, respectively.{ The signal modes are the nearest lower-neighbour WG modes to the `pump' WG mode, with no other high-$Q$ WG mode existing between them.}


\begin{figure}[b!]
	\centering
		\includegraphics[width=0.45\textwidth]{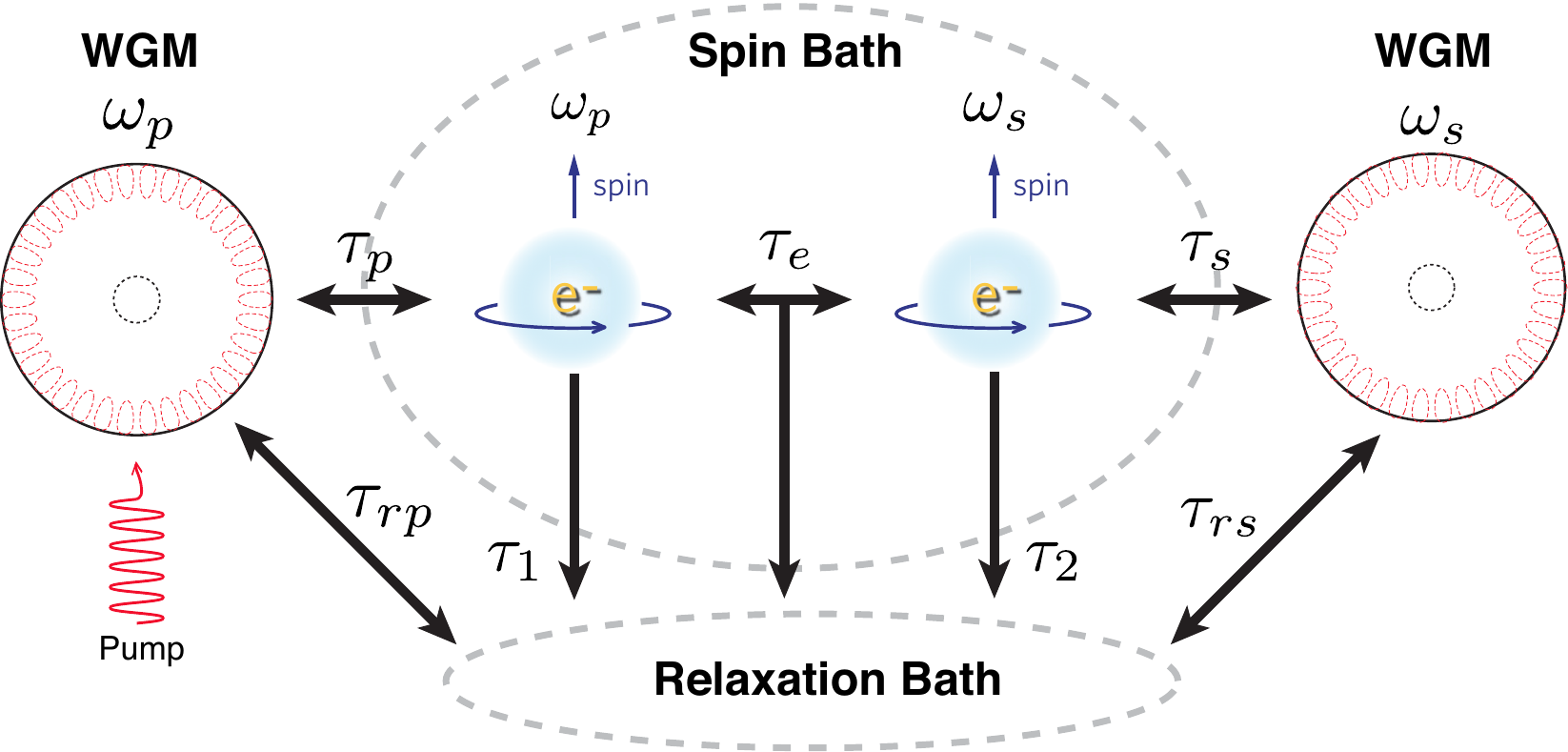}
	\caption{The spin bath system model demonstrates WG mode-bath ($\tau_p$ and $\tau_s$), spin-spin ($\tau_e$) couplings as well as direct dissipation processes ($\tau_1$, $\tau_2$, $\tau_{rp}$ and $\tau_{rs}$)}
	\label{fig:2oscillators}
\end{figure}


In the crystal under study, both the pump and signal WG modes lie within the bandwidth of the broadened Fe$^{3+}$ ESR. Thus, they are both effectively coupled to the same bath of Two Level Systems (TLS) as described in Fig. \ref{fig:2oscillators}.
The scheme consists of four subsystems: the pump WG mode, signal WG mode, the broadened spin bath of weakly interacting two-level systems (TLS), and the external environment or `relaxation bath'. The ensemble of TLS is physically present due to dilute Fe$^{3+}$ electron spin energy levels, in particular $\Ket{1/2}$ and $\Ket{3/2}$ spin angular momentum states. None of the other spin states are involved in the experiment.

The system Hamiltonian can be written as follows:	

\begin{equation}
\left. \begin{array}{rl}
H=\hbar\sum_x \{\omega_x q_x^\dagger q_x +\sum_{j=1}^N g_{xj}(\sigma^{+}_{j}q_x+q^{\dagger}_x\sigma^{-}_{j})\} \\
+\frac{\hbar}{2}\sum_{j=1}^N\omega_j\sigma_j^z +H_{int}+H_{ef}, 
\end{array} \right. 
\label{ham}
\end{equation}	
where $x\in\{p,s\}$ describes either the pump (denoted $p$) or signal (denoted $s$) mode, $q_x$ is either $a$ or $b$, representing the creation and annihilation operators for a photon in the pump and signal modes respectively, $\sigma_j^z = \ket{3/2_j}\bra{3/2_j}-\ket{1/2_j}\bra{1/2_j}$ is the atomic inversion operator for the $j^{\text{th}}$ TLS, $\sigma^{+}_{j}=\ket{3/2_j}\bra{1/2_j}$, $\sigma^{-}_{j}=\ket{1/2_j}\bra{3/2_j}$ are the excitation and annihilation operators for the $j^{\text{th}}$ TLS, and $\omega_j$ is the corresponding level splitting. $H_{int}$ and $H_{ef}$ are spin-spin and spin-lattice interaction terms.

The first summation in Eq. \ref{ham} describes two WG modes coupled to the common spin bath (shown as $\tau_p$ and $\tau_s$ in Fig. \ref{fig:2oscillators}), whereas the second term describes a bath of TLS with an inhomogeneous distribution of energy level splittings $\omega_j$.
The interaction term $H_{int}$ of the Hamiltonian describes the exchange of energy between different spins within the bath (labelled $\tau_e$ in Fig.~\ref{fig:2oscillators}), with a consecutive dissipation of the energy difference into the relaxation bath:
 \begin{equation}
H_{int}=\sum_{j< j^\prime} g_{sjj^\prime} \sigma^{-}_{j}\sigma^{+}_{j^\prime}e^{i\Delta\omega t}+ \text{h.c.},
\label{inter}
\end{equation}
where $\Delta\omega=\omega_j-\omega_{j^\prime}$, and $j$ and $j^\prime$ run over all spins of the ensemble. 
 The Hamiltonian that describes the coupling of each mode to the environment through the spin bath ($\tau_1$ and $\tau_2$ processes) is given by $H_{ef}=\sum_{j=1}^Ng_{lj}\sigma^{-}_{j}f^{\dagger} + c.c.$ where $f^{\dagger}$ is a creation operator for a mode in the relaxation bath system. This term is responsible for nonlinear losses of the WG modes. The $\tau_{rx}$ processes (Fig. \ref{fig:2oscillators}) have been neglected due to the low losses from WG modes compared to the spin bath.

Unlike in a conventional maser system, the ions of the bath here are effectively coupled. This interaction is attributed to strong spin-lattice interaction of dilute Fe$^{3+}$ ions in sapphire\cite{KorPro} described by the so-called spin-lattice anharmonicity term. This phonon mediated interaction is more probable than a direct magnetic spin-spin interaction due to the strong spin-lattice coupling, as evidenced by a number of other effects such as a considerable broadening of the Fe$^{3+}$ ESR. The broadening of the ion transitions, which is measured to have a full width at half maximum of 27~MHz \cite{symmons}, confirms that there is a strong coupling of the ions to the crystal lattice.
In addition, ESR spectroscopy measurements of the crystal under study\cite{warrick} confirm that Fe$^{3+}$ is the only ion present with an electron spin resonance in the frequency range studied in the present case (12 GHz), so no other available energy levels exist for the individual ions.

\begin{figure}[t!]
		\includegraphics[width=0.45\textwidth]{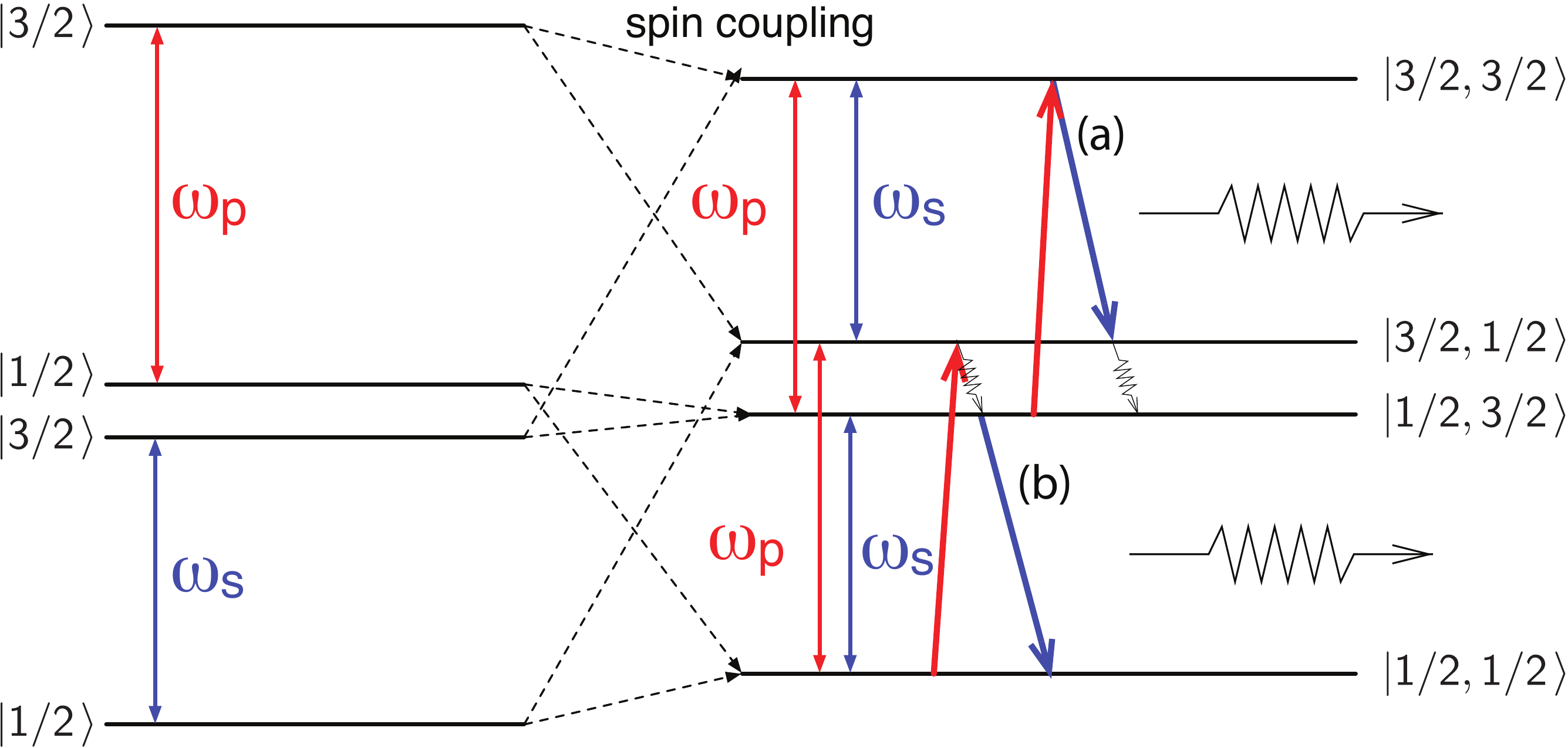}
		\caption{Energy level diagram of the uncoupled TLS (left) and coupled spins (right) in the weak coupling limit. 
		The resulting four level system exhibits two indistinguishable types of processes of population inversion and consequent photon emission.}
		\label{fig:coupledspin} 
\end{figure}

\begin{figure}[h!]
	\centering
	\includegraphics[width=0.4\textwidth]{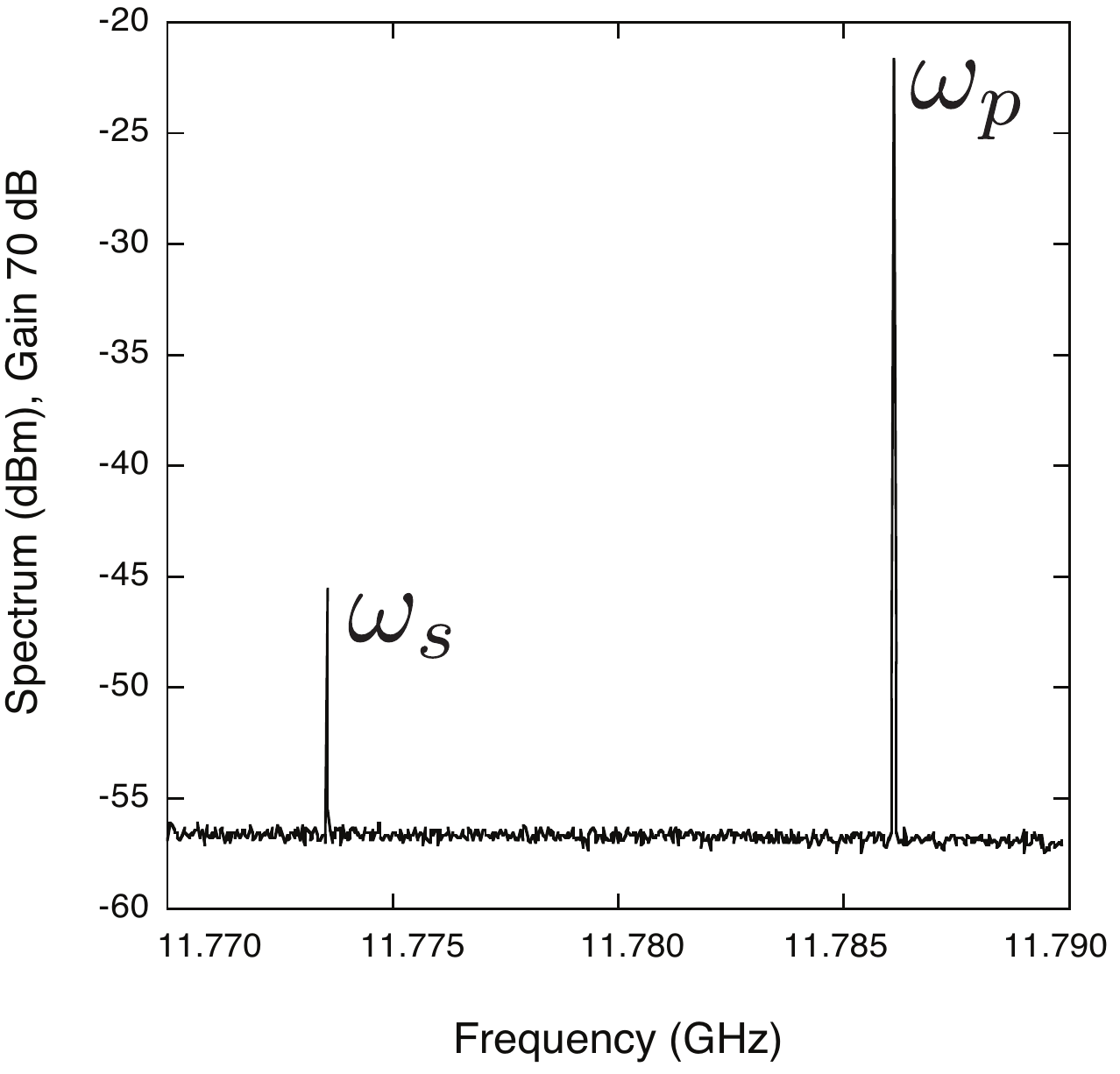}
	\caption{Spectrum analyser reading showing microwave emission at the signal WG mode frequencies when the pump mode at 12.037493 GHz has achieved the generation threshold}
	\label{fig:spectrum}
	\end{figure}

The spin-spin interaction term (Eq.~\ref{inter}) results in a situation where it becomes impossible to treat each ion independently, as is done in standard maser systems where the ion states in the bath are completely separable. In the present system, all the elements (TLS) of the bath have to be considered together to give the system state. As a consequence of this, the bath must be described as a whole, inseparable system with multiple energy levels. This can be done by introducing collective atomic operators\cite{ritsch} for each narrow frequency range of ESR where $\omega_i$ could be approximated to be the same, and then diagonalising the resulting approximate Hamiltonian of coupled Harmonic Oscillators. 
 Regardless of this complexity, the structure of the WG modes makes it possible to probe only two transitions of the bath ($\omega_s$ and $\omega_p$) of the collective system. This implies that only TLS with $\omega_j$ within the WG mode bandwidths near $\omega_s$ and $\omega_p$ can be coupled to. Effectively, this manifests as a four-level system (see Fig. \ref{fig:coupledspin}) with only two transitions $\omega_s$ and $\omega_p$.  Thus, the whole ensemble of interacting, inhomogeneously broadened spins (ion impurities) is seen by the pump and signal WG modes as an ensemble of effective three-level systems. This is because, energetically, only one type of process is possible (both the processes labeled (a) and (b)  in Fig.~\ref{fig:coupledspin} are equivalent). So, pumping above a certain threshold results in the system exhibiting masing (Fig.~\ref{fig:spectrum}) due to transitions between the corresponding energy levels (Fig.~\ref{fig:coupledspin})). The experimental observation of this net microwave emission from the solid-state WG mode cavity is shown in Fig.~\ref{fig:spectrum}.

When the crystal is pumped at 12.04 GHz, corresponding to the ``pump" WG mode ($\omega_p$), which lies at the centre frequency of a broadened Fe$^{3+}$ ESR, the generation of a signal ($\omega_s$ in Fig.~\ref{fig:spectrum}) located at one of two lower frequency WG modes (11.773 GHz or 11.786 GHz) is observed. The pump and signal WG modes are coupled through a broadened spin-bath (Fe$^{3+}$ $\ket{1/2}\rightarrow\ket{3/2}$ spin transition) as described above. 

ESR of Fe$^{3+}$ in sapphire has been measured previously \cite{karim2} to have a minimum range of influence of 189~MHz. {These results are confirmed for the crystal under study through numerous observations including power and magnetic field dependence of the WG mode resonance frequencies. Such dependencies can only arise through the interaction of WG modes with an ensemble of spins.}.

\begin{figure}[t!]
	\includegraphics[width=0.45\textwidth]{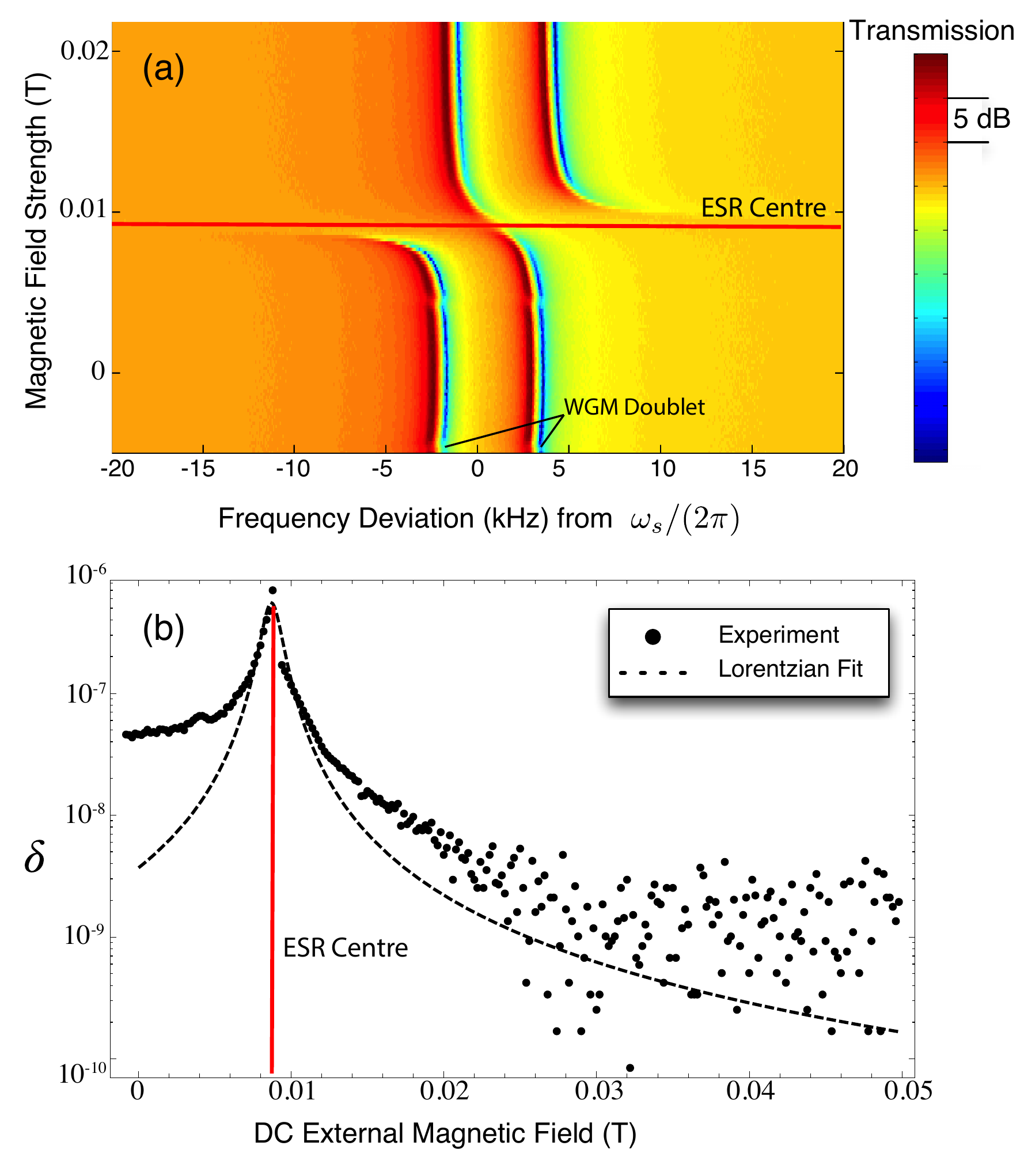}
	\caption{(a) Spectroscopy results (transmission through the cavity as a function of the external magnetic field and detuning frequency) showing frequency shift of the lower $\omega_s$ WGM (a doublet) caused by coupling to a transition of Fe$^{3+}$ ion. {The dark red regions correspond to the peak of the resonance, and hence the WG mode's frequency location for a given magnetic field. (b) Fractional frequency deviation (dots) of the lower $\omega_s$ WG mode lower doublet from its resonance frequency far from the ESR. Lorenzian line shape (dashed curve) fits the results within a narrow magnetic field range within the vicinity of the peak of the ESR.}}
	\label{fig:ESR}
\end{figure}

{ Firstly, the external magnetic field sensitivity can be used as a measure of spin-photon coupling. The application of an external DC magnetic field causes observable frequency shifts of WG modes due to their coupling to spins exhibiting the Zeeman effect. The maximum frequency deviation is achieved when the ESR centre frequency coincides with the WG mode. While for the pump mode this coupling is evident since its resonant frequency corresponds to the zero field splitting of the $\ket{1/2}\rightarrow\ket{3/2}$ transition, the coupling between the signal mode and the ensemble at zero field has been measured using a spectroscopic approach. 
Fig.~\ref{fig:ESR}(a) demonstrates the transmission through the cavity near $\omega_s$ as a function of the external magnetic field. The interaction between the ESR and WG mode is evident at 87.2 mT, corresponding to a frequency offset of 244 MHz, which is the difference between the $\omega_p$ and $\omega_s$ frequencies. Although even at zero field, $\omega_s$ is deviated from its reference value at infinity where the effect of the ESR is negligible. To demonstrate this, the frequency deviation is calculated using:
 \begin{equation}
\delta = \Big|\frac{\omega_s(B)-\omega_s(B\rightarrow\infty)}{\omega_s(B\rightarrow\infty)}\Big|,
\label{dev}
\end{equation}
and is shown in Fig.~\ref{fig:ESR} (b). This result shows that the leading edge of the spin transition is indeed present for the $\omega_s$ WG mode at zero field, affirming the assertion that the two WGMs ($\omega_p$ and $\omega_s$) are coupled. It also demonstrates that the leading edge of the ESR line shape is significantly different from predictions made by the Lorentzian line-shape fit made within a narrow range of fields near the ESR interaction. While the fitted Lorentzian curve predicts up to a 19 dB drop in the ESR coupling, experimental results give a factor of $\sim$15, calculated as a ratio between frequency deviations at zero field and at the interaction. This means that the number of spins participating in the interaction with the signal mode $\omega_s$ is only 15 times lower than that interacting with the pump mode. The Lorentzian fit gives an ESR bandwidth of 28.7 MHz (0.715 mT), in good agreement with previously reported results \cite{symmons}. The reason for significant divergence from the Lorentzian shape could lay in spin-spin interaction within the TLS bath. Note that Fig.~\ref{fig:ESR}(b) is symmetric about zero magnetic field.}

\begin{figure}[h!]
	\includegraphics[width=0.4\textwidth]{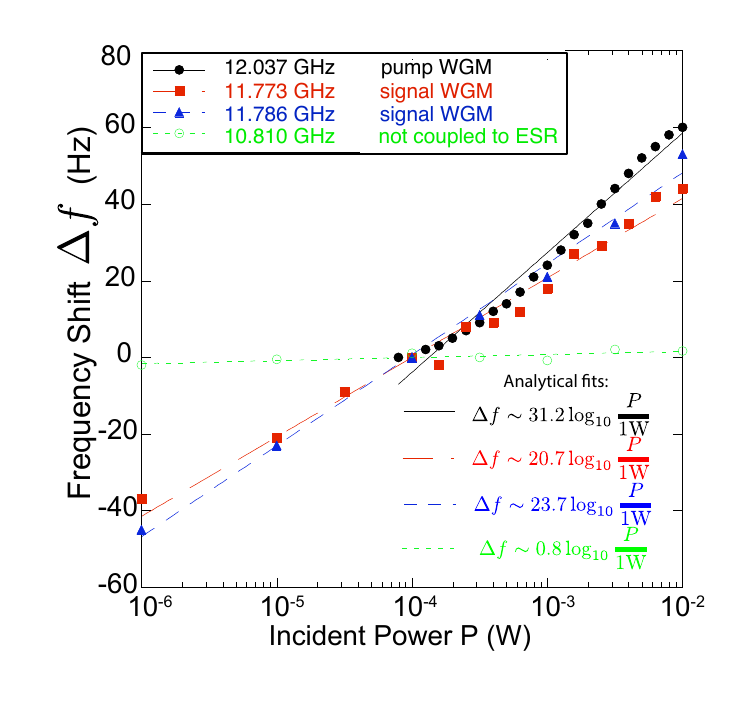}
	\caption{Incident power dependence of resonance frequencies of some WG modes. While three modes participating in spin bath maser demonstrate considerable frequency shift and thus coupling to the same spin ensemble, a mode out of the ESR range is power independent. }
	\label{fig:fshift}
\end{figure}

Secondly, {while photon modes in the pure sapphire crystals do not demonstrate any significant power dependence up to very high pumping powers, mode coupling to an ensemble of TLS causes strong nonlinear effects that are related to the saturation of the ensemble. As spins move from their normal to saturated states, they alter the microwave susceptibility of the crystal induced by their presence, thus causing a frequency shift in the WG modes supported in the resonator. Thus, by identifying the nonlinear behaviour of a particular WG mode, e.g. by frequency sensitivity to incident power, it is possible to determine its coupling to an ESR.
In this way, frequency shifts of the pump mode and both signal WG modes were observed (see Fig. \ref{fig:fshift}) as the input pump power was varied. For comparison, another WG mode at 10.8 GHz ($\sim$1.24 GHz offset from the centre of the ESR) was characterised, which demonstrates no frequency shift with increasing input power as it is outside the range of influence of the impurity ions. Since only the two closest lower frequency WG modes are effectively coupled to the same ESR as the pumping WG mode, only two signal frequencies could be obtained in the masing process.}


\begin{figure}[t!]
	\centering
			\includegraphics[width=0.4\textwidth]{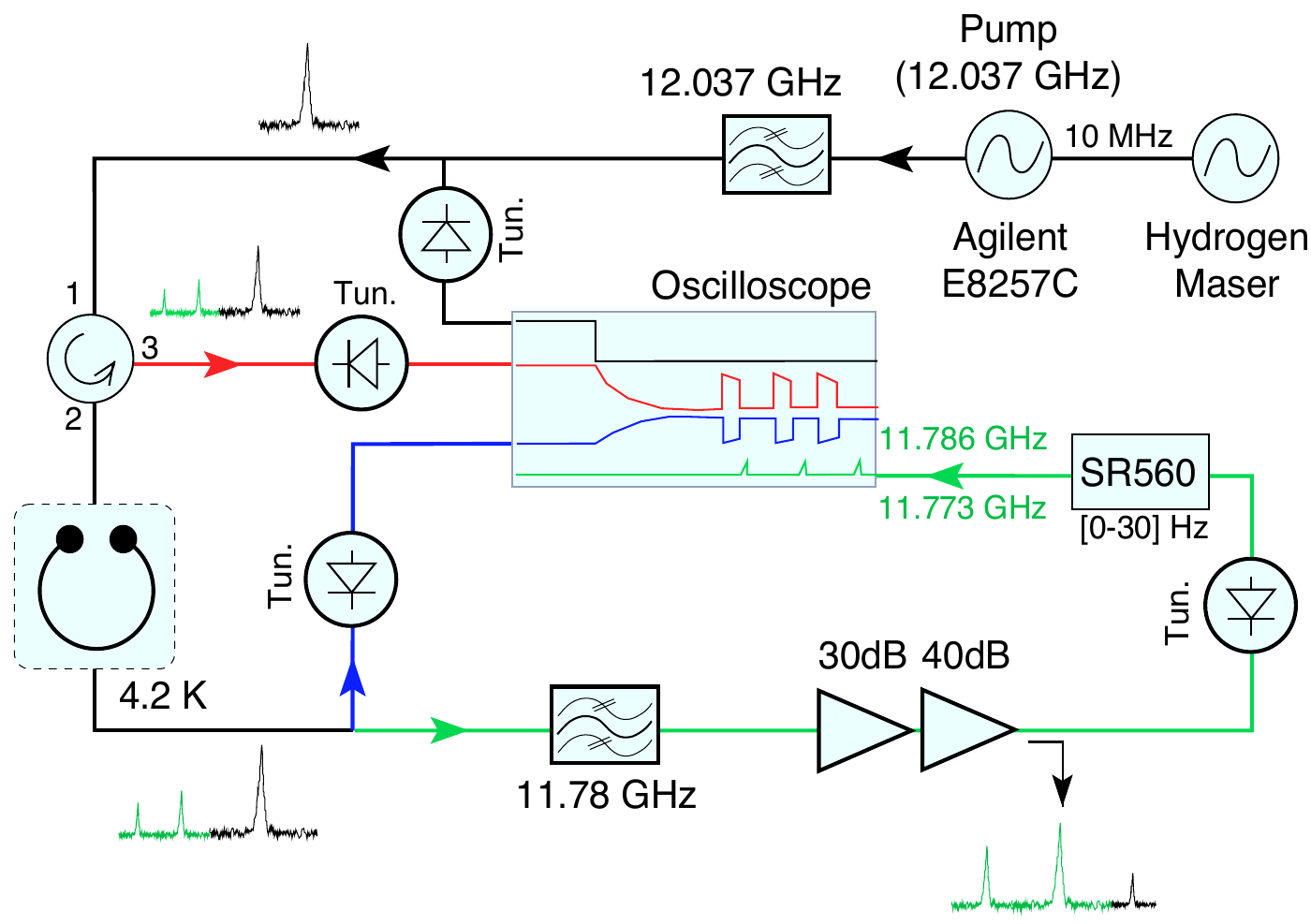}
	\caption{Experimental set-up}
	\label{fig:experiment}
\end{figure}

The signal mode frequencies were generated by continuously pumping the pump WG mode at high power ($\sim$23 dBm). The experimental setup is shown in Fig. \ref{fig:experiment}. The pump signal at $\omega_p$ was applied to the crystal using an Agilent E8257C synthesiser, referenced to a commercial hydrogen maser. The reflected and transmitted signals, as well as a filtered and amplified portion of the reflected signal at 11.77 GHz, were then observed with a digital oscilloscope. The combined gain of the amplifiers used was 70 dB. The filtered and amplified portion was simultaneously observed on an Agilent E4448A spectrum analyser. \\



\begin{figure}[t]
			\centering
			\includegraphics[width=0.4\textwidth]{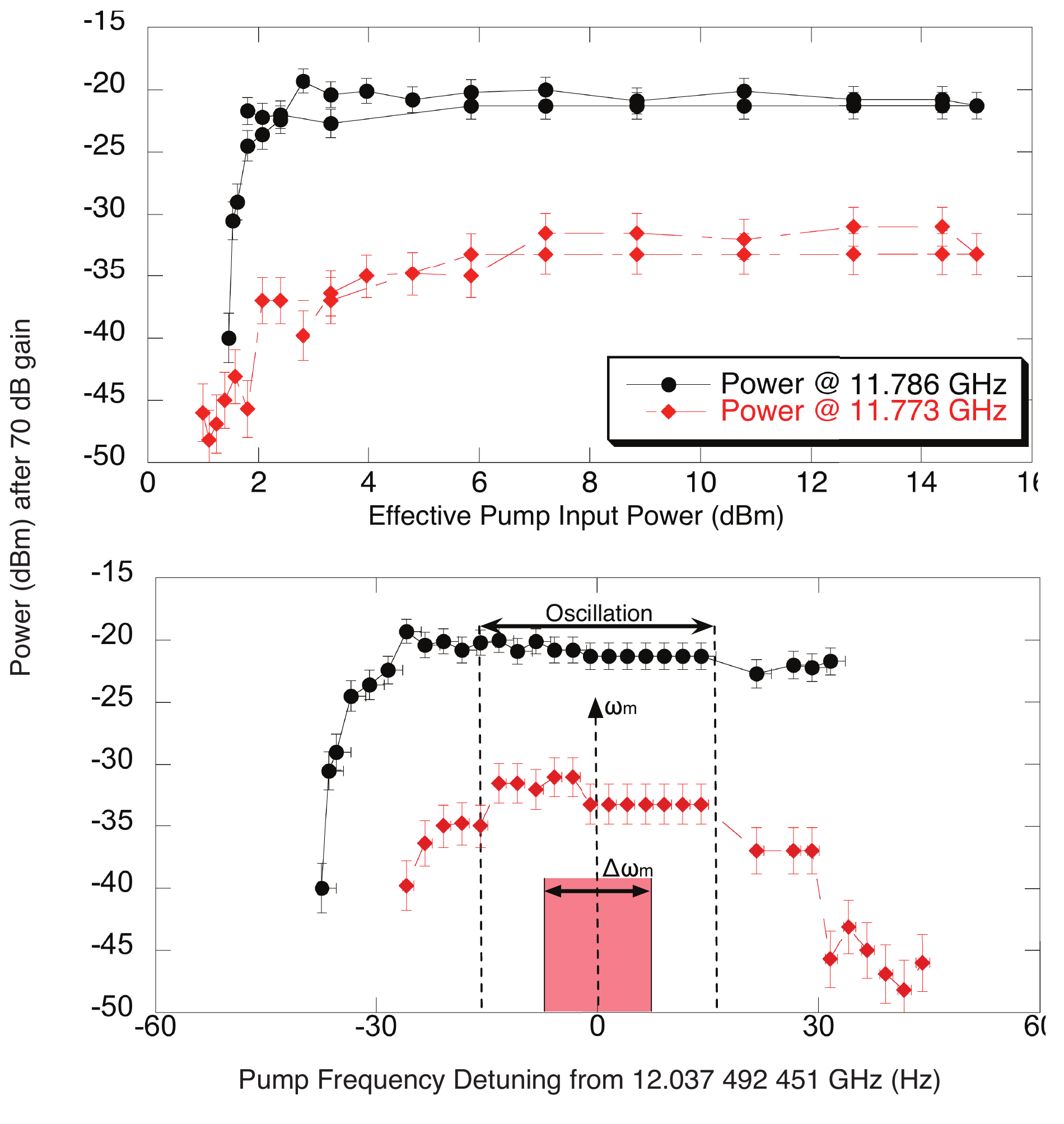}
			\caption{Output power of the signal and pump modes as a function of incident power (top) and pump frequency detuning from the lower doublet of the pump WG mode $\omega_p$ (bottom). The thresholds are clearly observed at $2.3$dBm and $\sim\pm40$Hz. }
			\label{fig:sweep}
\end{figure}

\indent The signal output from the ``signal'' WG mode was seen to operate in two regimes: continuous maser operation and a quadrature self-oscillation regime, depending on the frequency detuning and the corresponding injected power, shown in Fig.~\ref{fig:sweep}. The continuous signals (in the non-oscillating regime of Fig.~\ref{fig:sweep}) are effectively generated by the spin-bath solid-state system, as described by Fig.~\ref{fig:coupledspin}. The input pump signal excites the pump WG mode ($a^\dagger a$ in Eq.~\ref{ham}), which in turn excites the bath transitions corresponding to this frequency into their higher energy state ($\omega_p$ transition in Fig.~\ref{fig:coupledspin}.) due to the interaction given by the field-ion coupling term in Eq.~\ref{ham}. This creates a situation of population inversion. Due to the broadening of the Fe$^{3+}$ spin bath, these spins are coupled to a number of other WG modes located within the ESR bandwidth ($H_{int}$ given by Eq.~\ref{inter}). The subsequent relaxation of the bath can generate a photon with energy $\omega_s$ corresponding to the nearest lower-neighbour WG mode. Since the crystal confines photons of this energy, the generated photon leads to the stimulated emission of other photons of the same energy. In such a way, the signal WG mode, due to its extremely high quality factor, enhances the stimulated emission from the bath in its corresponding narrow frequency band.
 As the detuning is decreased and the corresponding power injected into the pump mode is increased, the threshold for microwave signal generation is reached \cite{karimthesis}, and the $\omega_s$ signals are observed to appear. This is the threshold corresponding to the continuous masing regime. Although the detuning frequency is swept, it is the injected WG mode power that exhibits the generation threshold. 
 Further tuning approaching resonance allows more photons to enter the system at $\omega_p$.  

\begin{figure}[t]
			\centering
			\includegraphics[width=0.4\textwidth]{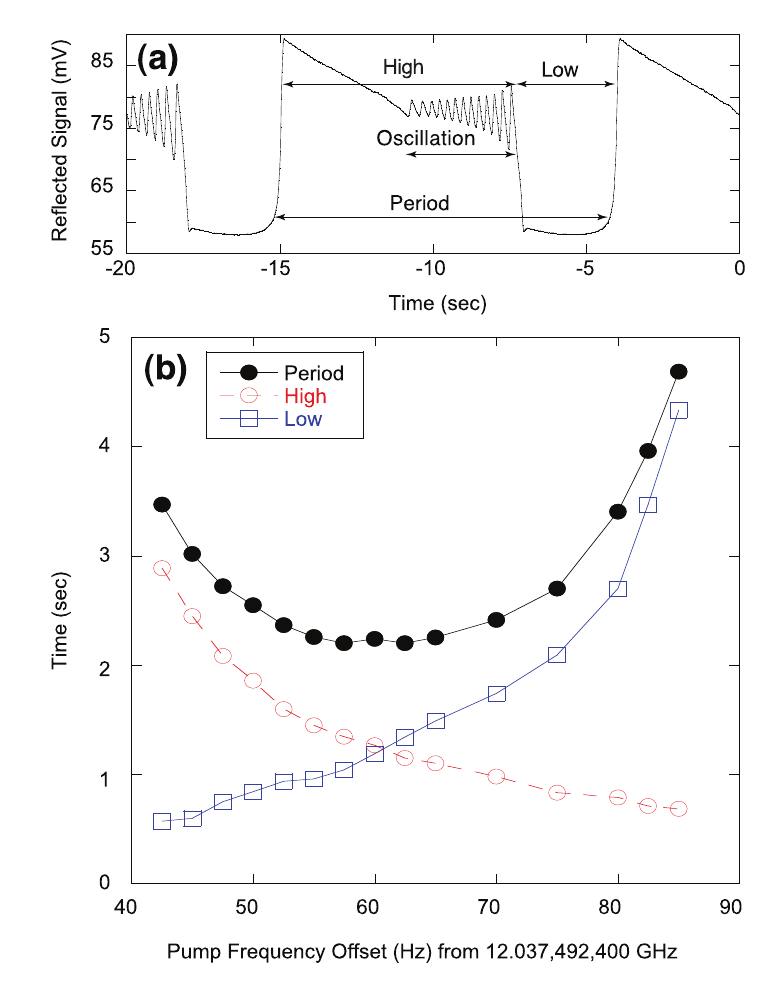}
			\caption{(a) Oscillogram of the reflected signal with a continuous, constant pump input. (b) shows the dependence on pump frequency of the oscillation{'}s period for the lower WGH$_{17,0,0}$ mode doublet.}
			\label{fig:VDP}
\end{figure}

Above another threshold in both incident power and detuning, the normally stable magnitudes of the signal and pump began to oscillate as a function of time. This phenomenon occurs over the range of frequencies outlined in Fig.~\ref{fig:sweep} and takes the temporal form of Fig.~\ref{fig:VDP}(a). This is the quadrature oscillation regime which results from the fact that the signal mode operates at the ion saturation threshold resulting in strong nonlinearity in the loss term of the corresponding equation of motion. The corresponding nonlinearity\cite{karim1, karimthesis} with the associated degrees of freedom of the WG mode results in a system of two coupled Van der Pol oscillators in the signal quadrature space. Such a system has been derived analytically from the physical description of the WG mode resonator.
 The experimentally observed relaxation oscillation behaviour is indeed very well described by a Van der Pol oscillator model\cite{VDP}. It is important to note that at least two coupled WG modes are needed for this type of quadrature oscillation behaviour due to a lack of sufficient degrees of freedom for only one mode.\\

In summary, our results have demonstrated the possibility of a novel maser mechanism in a solid-state system. 
The effect is achieved in a bath of weakly interacting, inhomogeneously broadened TLS which constitute an inseparable spin bath. The pump signal creates an effective population inversion in the coupled spin system, and high-$Q$ WG modes of the crystal considerably increase the photon lifetime. This leads to an enhanced probability of photon-bath interaction, and thus stimulated emission at the corresponding resonant frequencies.
The net microwave emission at the signal frequency exhibits power and detuning frequency thresholds. 
The nature of the effect has been demonstrated to be dependent on the pump detuning from resonance, and can therefore be controlled, and occurs in both a continuous output and an oscillating regime.
 
\section*{Acknowledgements}
This work was supported by Australian Research Council grants CE110001013 and FL0992016.

\section*{References}

%

\end{document}